\documentclass[lettersize]{article}
\usepackage{amsmath,amsfonts}
\usepackage{algorithmic}
\usepackage{array}
\usepackage[caption=false,font=normalsize,labelfont=sf,textfont=sf]{subfig}
\usepackage{textcomp}
\usepackage{stfloats}
\usepackage{url}
\usepackage{verbatim}
\usepackage{graphicx}
\usepackage{cite}

\usepackage[T1]{fontenc}  % Use T1 encoding, which supports Polish characters
\usepackage[utf8]{inputenc}
\usepackage{multicol} % Required for creating multiple columns in slides
\usepackage{float} % Allows putting an [H] in \begin{figure} to specify the exact location of the figure
\usepackage{wrapfig} % Allows in-line images such as the example fish picture
\graphicspath{{./images/}}		%chemin vers les images

\usepackage{booktabs} % Required for better table rules : \toprule, \midrule, \bottomrule and set the width of the columns
\usepackage{multirow} % Required for multirows in tables

\usepackage{amsmath} % Required for some math elements
\usepackage{amsthm} % to write theorems, definitions, lemmas, ...
\usepackage{amssymb} % more math symbols

\usepackage{hyperref} % manage hyperlinks
\usepackage{xcolor}

\newtheorem{theorem}{Theorem}[section]
\theoremstyle{definition}
\newtheorem{definition}[theorem]{Definition}

\begin{document}

\date{}
\title{Fast~and~Robust~Flocking of~Protesters~in~Street~Networks}
\author{Guillaume Moinard,\footnote{Contact author: \url{guillaume.moinard@lip6.fr}} and Matthieu Latapy\\
Sorbonne Université, CNRS, LIP6, F-75005 Paris, France}
\maketitle

\begin{abstract}

We present a simple model of protesters scattered throughout a city who want to gather into large and mobile groups. This model relies on random walkers on a street network that follow tactics built from a set of basic rules. Our goal is to identify the most important rules for fast and robust flocking of walkers. We explore a wide set of tactics and show the central importance of a specific rule based on alignment. Other rules alone perform poorly, but our experiments show that combining alignment with them enhances flocking, and that obtained groups are then remarkably robust.

\end{abstract}

\section{Introduction}

Consider the following scenario. Protesters are scattered throughout a city and want to gather into groups large enough to perform significant actions. They face forces that may break up groups, block some places or streets, seize communication devices carried by protesters, or even shut down or compromise communication systems. As a consequence, protesters only have access to local information on people and streets around them. Furthermore, formed protester groups must keep moving to avoid containment by adversary forces and perform actions.

\begin{figure}[!h]
    \centering
    \includegraphics[width=\linewidth]{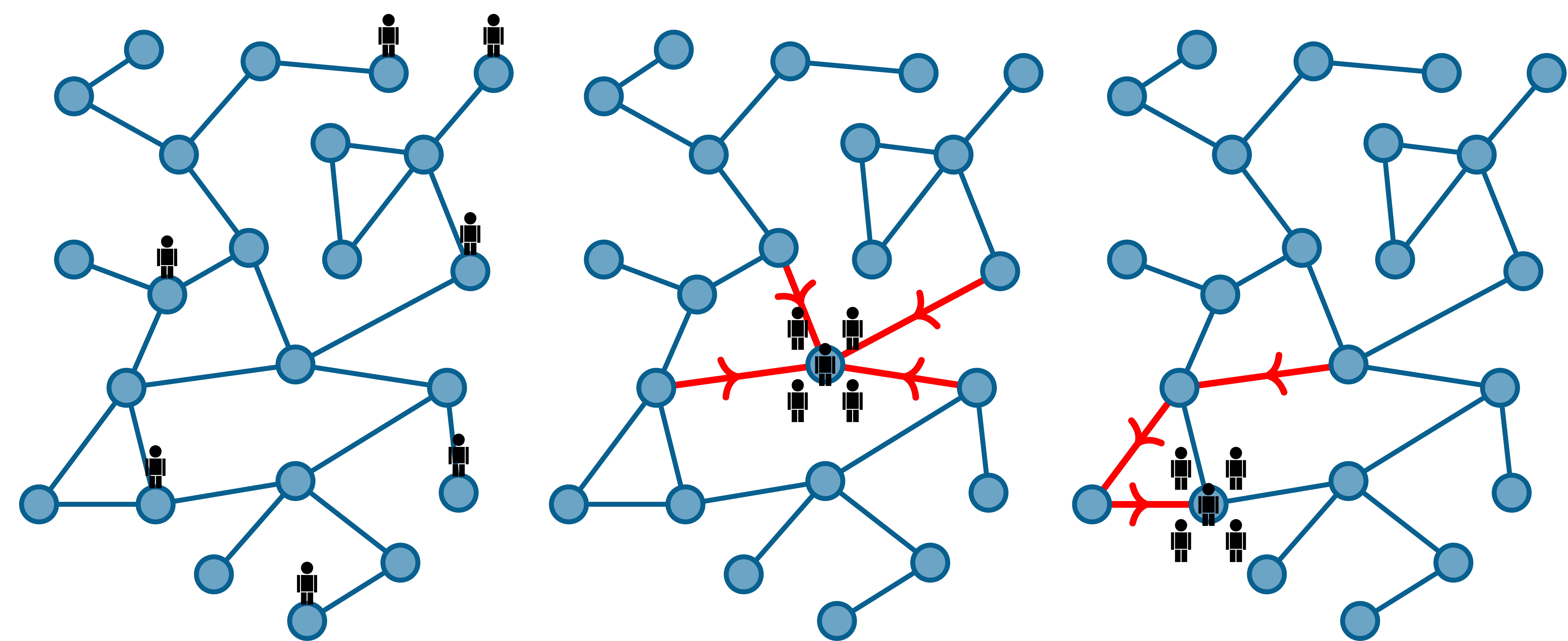}
    \caption{We expect walkers scattered on a street network (left) to quickly gather (center) and then flock (right).}
    \label{fig:tikz}
\end{figure}

In this scenario, protesters need a distributed and as simple as possible protocol, that uses local information only and ensures the fast formation of significantly large, mobile, and robust groups.
We illustrate these objectives in Figure~\ref{fig:tikz}, and we call the expected collective behavior \emph{flocking}.
Our goal in this paper is to identify the key building blocks for flocking.

To do so, we first model the city as a network of streets and intersections, and we rely on deliberately strong simplifications to model protesters as biased random walkers on this network.
We then define a set of basic rules that walkers may use as building blocks to compose \emph{tactics}, as well as metrics that quantify key group features. We perform extensive experiments in order to explore how various tactics behave regarding these metrics. We also explore group robustness when adversary forces break them up while they follow an effective tactic.
Finally, we discuss related work and highlight the originality of our contribution.

%An extended abstract of this work was presented at ASONAM 2024, see~\cite{moinard2024fast}. We propose here a more in-depth analysis of tactics and their efficiency. In particular, we detail the behavior of each rule in a single street, which gives interesting insight on the overall behavior of tactics. We also add the important study of tactic robustness.

\section{Framework}
\label{sec:framework}

This section presents our simulation framework. It models cities as undirected graphs that we call street networks. Then, protesters are biased random walkers on this network. They move from node to node following simple rules that we detail in next section. We keep our model as simple and minimal as possible in order to make it easy to explore the key principles behind flocking; our goal is {\em not} to model real-world protester behaviors, which is another important and challenging topic.

\subsection{Street networks}
\label{sec:streets}

In order to model real-world cities, we leverage OpenStreetMap data and the OSMnx library~\cite{boeing2024modeling}. For a given city, we use this library with its default settings to extract the graph $G=(V,E)$ defined as follows: the nodes in $V$ represent street intersections in this city and the links in $E \subseteq V\times V$ represent the pieces of streets that exist between them. We focus on drivable streets because protesters generaly target them in order to maximize their impact on city routine. The graph $G$ is undirected, meaning there is no distinction between $(u,v)$ and $(v,u)$ in $E$. In addition, we denote by $N(v) = \{u, (u,v) \in E\}$ the set of neighbors of any node $v$ in $V$.% and by $N^+(v)$ the set $N(v) \cup \{v\}$.

We performed experiments on a wide set of large worldwide cities of diverse sizes and structures. This led to no significant difference on obtained results. We therefore use a typical instance, namely Paris, to present our work in this paper. This street network has 9\,602 nodes, 14\,974 links, leading to average degree 3.1. Its diameter is 83 hops and its average distance is 39.4 hops\,\footnote{The distance between two nodes is the length of a shortest path between them. The average distance is the average of this length over all pairs pairs of nodes. The diameter is the maximal distance.}. The average street length is 99 meters, and the average walking distance is 5\,552 meters.

\subsection{Walkers}
\label{sec:walkers}

Given a network $G=(V,E)$, we consider a set $W$ of walkers numbered from $1$ to $|W|$. We denote the location of walker $i$ at time $t$ by $x_i(t) = v$, with $v\in V$. At each time step $t$, walker $i$ moves to node $x_i(t+1) \in N(v)$. Rules we present in the following sections determine the choice of $x_i(t+1)$.

For any link $(u,v)$ in $E$, we also define the {\em flux} of walkers from $u$ to $v$ as $J_{u\rightarrow v}(t) = |\{i, x_i(t-1)=u, x_i(t)=v\}|$.
We call {\em group} the set of walkers at a given node $v$ at a given time $t$: $g_v(t) = \{i, x_i(t)=v \}$. We denote by $n_v(t) = |g_v(t)|$ the number of walkers located at node $v$ at time $t$. We denote by $g(t) = |\{g_v(t), v\in V, g_v(t)\ne \emptyset\}|$ the number of non-empty groups at step $t$.

This definition allows groups of only one walker. We will see in Section~\ref{sec:metrics} how to detect and avoid tactics that lead to such uninteresting groups.

\subsection{Discretization}
\label{sec:discr}
\begin{figure}[h!]
  \begin{center}
      \includegraphics[width=\linewidth, trim=0 0 0 100,clip]{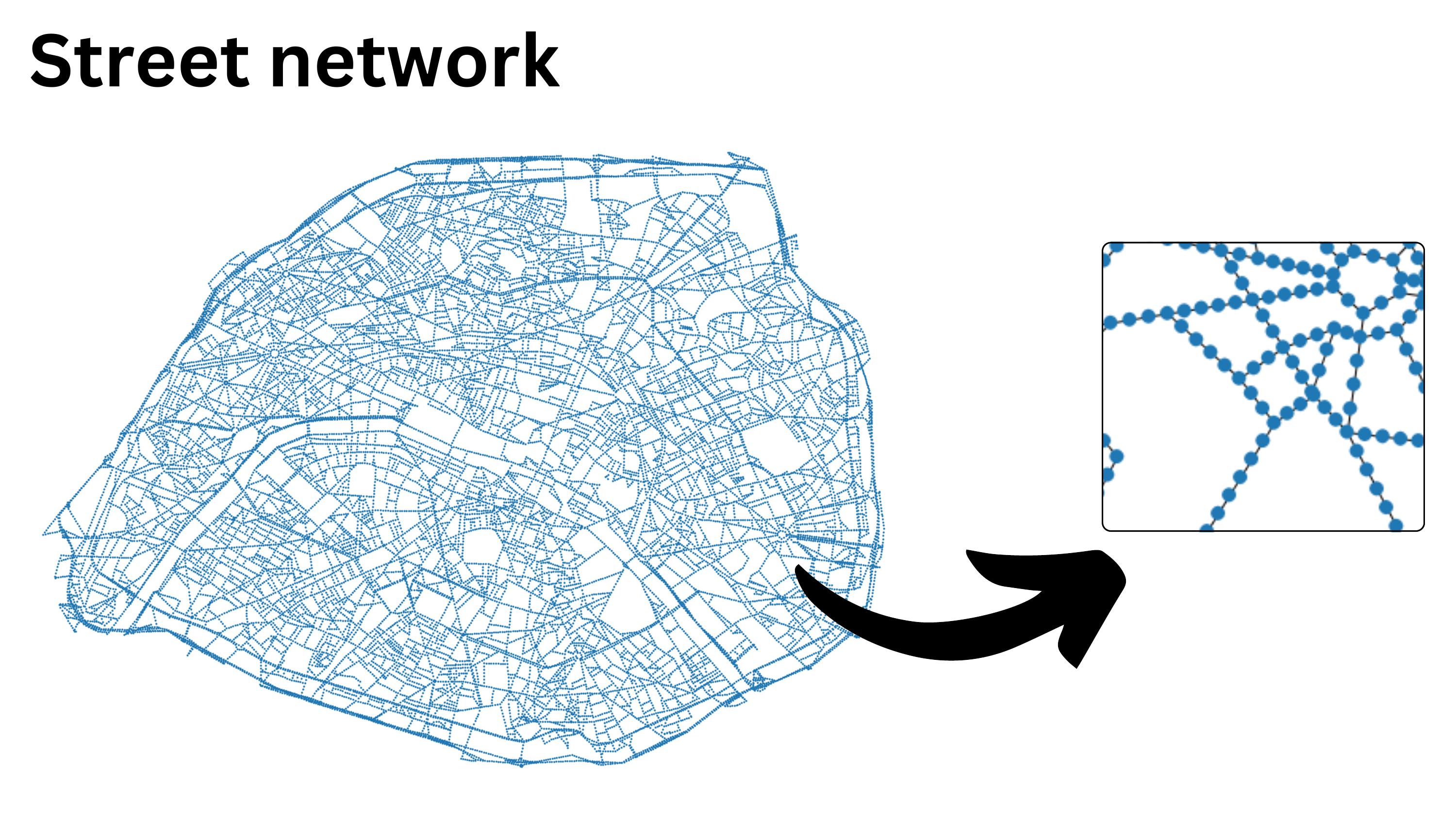}
    \caption{A piece of the discretized street network around {\em Place de la Nation} in Paris.}
    \label{fig:paris}
\end{center}
\end{figure}

The links of a street network generally represent street segments of very heterogeneous lengths with heavy-tailed distributions~\cite{masucci2009random}. Then, moving from an intersection to another one may have very different durations. In order to model this, we use a classical discretization procedure~\cite{neri2011totally} that we describe below. It consists in splitting each link of the street network into pieces connected by evenly spaced nodes. We illustrate this in Figure~\ref{fig:paris}.

More formally, given a link $(u,v)$ representing a piece of street of length $L$ in a street network and a discretization step $\delta$, we split each undirected link $(u,v)$ into $k$ new links, with $k = \lceil{\frac{L}{\delta}}\rceil$ as follows. We define a set of new nodes $V(u,v) = \{ w_0, w_1, ..., w_k \}$, the nodes from the original network being $w_0 = u$ and $w_k = v$. We also define a set of new links $E(u,v) = \{(w_i, w_{i+1}), i = 0, 1,...,k-1 \}$. Then we define the set of all nodes $V = \bigcup_{u,v} V(u,v) $ and the set of links $E = \bigcup_{u,v} E(u,v)$. %, where between the locations of these  nodes.
Finally, we model the street network by the graph $G = (V,E)$.

% \begin{figure}[h!]
% \centering
% \includegraphics[width=0.9\columnwidth]{images/nation_network_cropped.pdf}
% \caption{A piece of the discretize street network around {\em Place de la Nation} in Paris.}%, discretized with $\delta = 10$ meters.} %20 meters actually?
% \label{fig:paris}
% \end{figure}

% \begin{wrapfigure}[13]{r}{0.2\textwidth}
%   \begin{center}
%     \includegraphics[width=0.2\textwidth]{images/nation_network_cropped.pdf}
%     \caption{A piece of the discretized street network around {\em Place de la Nation} in Paris.}
%     \label{fig:paris}
% \end{center}
% \end{wrapfigure}

In the obtained graph, each link represents a street slice of length close to $\delta$. Then, the walkers defined in previous section consistently make a move of length approximately $\delta$ at each time step.
In this paper, we use $\delta$ equal to 10 meters, leading to a network of $n=130\,276$ nodes and $m=300\,736$ links for Paris. It gives a sufficient precision in our context, and experiments with other reasonable values of $\delta$ displayed no significant difference.

\section{Tactics}
\label{section:model}

At each time step $t$, each walker $i$ moves from location $x_i(t)$ to location $x_i(t+1)$ in $N(x_i(t))$. This section presents how we choose the new location $x_i(t+1)$.

\subsection{Available information}% for choices}

We assume that walkers have very limited capabilities. They have no long-term memory and no means of communication. They do not recognize other protesters and they do not have access to the location of other walkers, even on neighbor nodes. Instead, we assume they only have access to estimates of aggregated observables, such as the number of walkers on neighbor nodes and the flow intensity on surrounding links, as well as a very short-term memory.

More formally, walker $i$ may use the following information at time $t$:
\begin{itemize}
\item its previous location $x_i(t-1)$,
\item the number $n_v(t)$ of walkers on node $v$ for all $v$ in $N(x_i(t)) \cup \{x_i(t)\}$,
\item the flux of walkers arriving and leaving its current location $v=x_i(t)$, {\em i.e.} $J_{u\rightarrow v}(t)$ and $J_{v\rightarrow u}(t)$ for all $u$ in $N(v) $.
\end{itemize}

\newcommand{\bigomega}{\mbox{\scalebox{1.4}{$\omega$}}}

Thanks to the information above, each walker $i$ knows it previous location $u = x_i(t-1)$, its current location $v = x_i(t)$, and it has access to various criteria to decide its next location $w = x_i(t+1)$. Then, it needs a way to derive walking rules from criteria, and a way to combine these walking rules into a tactic.

\subsection{Criteria}

\noindent
A criterion is a parameter from which we construct walking rules. We consider the following set of criteria $\mathcal{C}$.

\begin{itemize}
\item {\em Random.} The walker makes no difference between all possible neighbors: the criterion has value $1$ for each of them.
\item {\em Propulsion.} The walker never goes back to its previous location $u$, and otherwise makes random moves: the criterion has value $0$ for $u$ and value $1$ for other neighbors of $v$.
\item {\em Attraction.} The walker preferably moves to nodes where there are already many walkers: the criterion is equal to the number of walkers $n_w(t)$ for each neighbor $w$ of $v$.
\item {\em Follow.} The walker preferably follows the most popular moves of other walkers: the criterion has value $J_{v\rightarrow w}(t)$ for each neighbor $w$ of $v$.
\item {\em Alignment.} The walker takes into account the net flux in both directions: the criterion is equal to $\Delta J_{v\rightarrow w}(t) = J_{v\rightarrow w}(t) - J_{w\rightarrow v}(t)$ for each neighbor $w$ of $v$.
\end{itemize}

Notice that the Alignment criterion may take negative values, when the flux towards $w$ is lower than the flux from $w$. This has a repulsion effect on walkers: they avoid rushing towards crowds that are already going to join them.

We chose these criteria because they are very simple and rely on local information only. They also model basic behaviors that may be observed in real-world protests, such as following others, avoiding going back, or joining already existing large groups. Last but not least, we also selected these criteria because they do not use any external information or communication, as wanted.

\subsection{Walking rules}

Let us consider a criterion $C_{u,v,w}(t)$. We define the corresponding walking rule using the classical \emph{logit rule}. It gives the
probability $\bigomega^C_{u,v,w}(t)$ that walker $i$ moves from $v$ to $w$, given the fact that it arrived at $v$ from $u$:
\begin{equation}
\label{eq:logit}
\bigomega^C_{u,v,w}(t) = \frac{e^{\beta \cdot C_{u,v,w}(t)}}{\sum_{z \in N(v)} e^{\beta \cdot C_{u,v,z}(t)}}
\end{equation}

Parameter $\beta \geq 0$ is the intensity of choice: it quantifies the influence of the criterion on walker choices.  If $\beta = 0$, the criterion has no influence and walkers make purely random choices. If $\beta \rightarrow \infty$, walkers necessarily choose a neighbor among the ones that maximize the criterion.

This logit equation is widely used in the literature because it has many convenient features~\cite{bouchaud2013crises}. In particular, it ensures $\bigomega$ is a positive and monotonic function of the criterion: the greater $C_{u,v,w}(t)$, the greater the chances to move to node $w$ at time $t$. It is also compatible with criteria having both positive and negative values, like the flux-based one above.

We conducted experiments with other functions with these needed features; they led to no significant differences.

\subsection{Rule aggregation}%Combination of rules}%

A \emph{tactic} is a linear combination of walking rules that defines the probability $\pi_{u,v,w}(t)$ to move from $v$ to one of its neighbors $w$ when coming from $u$:
\begin{equation}
\pi_{u,v,w}(t) = \sum_{C\in \mathcal{C}} \alpha_C \cdot \bigomega^C_{u,v,w}(t)
\label{eq:mastereq}
\end{equation}
where $\alpha_{C}$ is the coefficient of criterion $C$, with $\sum_{C\in \mathcal{C}} \alpha_C = 1$. Therefore, a tactic is defined by a set of criteria and their coefficients.

We call \emph{strict tactic} one that has all its coefficients, except one, set to zero and then always follows the same criterion.

In practice, at each time step, each walker selects a criterion $C$ with probability $\alpha_C$. Then, it computes its probability $\bigomega^C_{u,v,w}$ to go to each neighbor node $w$ and selects its new location accordingly. %{\color{red} We remark that some rules do not depend on $w$ and that we have $\pi_{u,v,w}(t) = \pi_{u,v}(t)$ for tactics based only on such rules.}

Notice that we may see a tactic as an inhomogeneous Markov chain, where transition probabilities are time-dependent. Still, at any given time, each walker makes a transition from its current state $(u,v)$ giving its previous location $u$ and its current location $v$, to its next state $(v,w)$ giving its current location $v$ and its next location $w$.

As transition probabilities at time $t$ depend on previous non-deterministic moves, formal analysis of such processes is generally out of reach~\cite{bremaud2001markov}. We therefore explore possible tactics using simulations.

\subsection{Baseline}
\label{section:baseline}

In addition to the tactics above, we consider a reference baseline that easily achieves flocking thanks to collective decisions. This means that, at each time step, all walkers at a given node make the same choice. We then obtain reference results that we expect our walker models to reach or outperform, even though they are unable to make collective decisions.

More formally, for each node $v$ at time $t$, a unique random neighbor $u$ of $v$ is chosen and all walkers $i$ such that $x_i(t)=v$ move to $u=x_i(t+1)$. This is equivalent to purely random walks of groups until they meet another group.

Since we consider non-bipartite connected graphs, it is well known that all groups will eventually merge. The number of needed steps is called the coalescence time \cite{Cooper2013}. Even though this number may be prohibitive, this means that the baseline successfully produces large groups. In addition, groups are mobile since, once formed, they perform purely random walks. As a consequence, this baseline makes a relevant reference for the success of flocking walkers.

%\input{theory}

%\subsection{Gathering and flocking metrics}
\subsection{Flocking metrics}
\label{sec:metrics}

In order to evaluate tactics, we need metrics describing how well walkers flock. We characterize \emph{flocking} as a gathering of walkers that keep exploring the network.

%Gathering is characterized by walkers located at the same node/neighbor at a given time~\cite{bhagat2022meet} and flocking is characterized by their ability to travel/explore the network.

This leads to the following \emph{score} definitions for a given run of a given tactic. %First, to measure the efficiency of a tactic for exploration, we define the coverage score.
First, we define the coverage score, a classical concept to measure the efficiency of a walk for exploration~\cite{masudaRandomWalksDiffusion2017a}.

%\begin{enumerate}
%    \item[\textbf{Def 1.}] The {\em coverage} $\mu_i(t)$ of walker $i$ is the number of distinct nodes a walker has already visited at time $t$: $\mu_i(t)= |\{v, \exists t' \leq t, x_i(t')=v\}|$. The {\em coverage score} $\mu(t)$ is the average walker coverage: $\mu(t)=\frac{1}{|W|}\sum_{i\in W} \mu_i(t)$.
%\end{enumerate}
%

\begin{definition}
The {\em coverage} $\kappa_i(t)$ of walker $i$ is the number of distinct nodes a walker has already visited at time $t$: $\kappa_i(t)= |\{v, \exists t' \leq t, x_i(t')=v\}|$. The {\em coverage score} $\kappa(t)$ is the average walker coverage: $\kappa(t)=\frac{1}{|W|}\sum_{i\in W} \kappa_i(t)$.
\end{definition}

Notice that the coverage score is monotonically non-decreasing with time and bounded by the number of nodes: $|V| \ge \kappa(t+1) \geq \kappa(t)$. In addition, if at time $t$ all walkers move to a node they already visited then $\kappa(t+1) = \kappa(t)$. In particular, walkers stuck in loops do not make the coverage score grow.

Second, we need a way to evaluate the ability of walkers to gather. The trivial approach consists in observing the average group size, a group being the set of walkers located at a same node. However, this makes no difference between two groups on neighbor nodes and two groups at very distant nodes. We therefore introduce the more advanced notion of {\em clusters}.
%, and the associated {\em gathering score}.
%for locally close groups and introduce the following score.  the gathering score must take into account that two neighbor groups demonstrate a greater gathering than two groups on distant nodes. The average size of groups is therefore not well suited. We define a core notion of

%\begin{enumerate}
%    \item[\textbf{Def 2.}] We call {\em cluster} a maximal connected sub-graph with walkers on all its nodes. The node-size and the walker-size of a cluster is the total number of nodes involved in the cluster, and the total number of walkers on these nodes, respectively.
%\end{enumerate}

\begin{definition}
We call {\em cluster} a maximal connected sub-graph with walkers on all its nodes. The node-size and the walker-size of a cluster is the total number of nodes involved in the cluster, and the total number of walkers on these nodes, respectively.
\end{definition}

If all walkers are in the same cluster then the average walker-size of clusters is $\gamma(t)=|W|$. If instead all walkers are in different clusters, then the average walker-size of clusters is $\gamma(t)=1$. However, if there is a unique walker at each node of the considered graph, then there is a unique cluster and it contains all walkers. Then, the average walker-size of clusters is maximal, despite the fact that there is actually no significant group and no gathering. This is why we need to consider both the walker- and node-size of clusters, which lead to the following definition of gathering and sprawling scores.

% \begin{definition}[gathering ratio $\gamma(t)$]
% The gathering ratio $\gamma(t)$ is the average number of walkers in non-empty groups: $\gamma(t) = \frac{|W|}{g(t)}$.
% \end{definition}
%When every agents are on different nodes, $\mathrm{g}(t)=|W|$. If instead all agents are on the same node, it equals $1$.

%\begin{definition}[cluster, gathering score $\gamma(t)$]
%We call cluster a maximal connected sub-graph with walkers on all its nodes. Then, the gathering score $\gamma(t)$ is the average number of walkers in clusters.
%\end{definition}

%\begin{enumerate}
%    \item[\textbf{Def 3.}]
%The {\em gathering score} $\gamma(t)$ is the average cluster walker-size at time $t$.
%The {\em sprawling score} $\sigma(t)$ is the average cluster node-size at time $t$.
%\end{enumerate}

\begin{definition}
The {\em gathering score} $\gamma(t)$ is the average cluster walker-size at time $t$.
The {\em sprawling score} $\sigma(t)$ is the average cluster node-size at time $t$.
\end{definition}

% \begin{definition}[cluster, sprawling ratio $\sigma(t)$]
% A cluster is a maximal connected sub-graph with walkers on all its nodes.
% The sprawling ratio $\sigma(t)$ is the average number of nodes of clusters.
% \end{definition}

%\begin{definition}[sprawling score $\sigma(t)$]
%The sprawling score $\sigma(t)$ is the average number of nodes in clusters.
%\end{definition}

Since our main goal is to obtain large groups, we are primarly interested in tactics with high gathering score. Secondarily, we are interested in high coverage score and low sprawling score. Indeed, high gathering scores together with low sprawling scores ensure that walkers form significant clusters: many walkers are grouped on a small set of close nodes. A high coverage score implies in addition that walkers continue to move in the network. Therefore, the combination of such scores ensures that walkers move in groups, which means that the tactic successfully achieves flocking.

%However, we need to ensure that clusters do not spread out too much. Indeed, a cluster with one walker on each node is not a good flocking pattern, even though it has a high gathering score.
%To verify such property, we introduce an additional metric.

%If the sprawling score is $1$, all groups are isolated from each other: whenever walkers are at a node, there is no walker on neighbor nodes. If instead the sprawling score is high, walkers form large clusters of neighbor groups. Its largest possible value is the total number of nodes in the network, meaning that there are walkers on each node.

%We are interested in tactics with low sprawling score, meaning that they succeed in merging neighbor groups.

\section{Exploration of tactics}
\label{section:measures}

We seek good tactics for our walkers. In particular, we seek tactics that produce flocking comparable to the baseline presented in previous section. Moreover, we want to obtain short time convergence, as our walkers model protesters that cannot walk forever. We therefore want tactics that produce flocking as fast as possible.

This section presents a wide set of simulations for realistic walk durations and explores which tactics perform best. We consistently provide interpretations of observed results and we deepen the study of the most effective tactics.

\subsection{Behaviors on a single street}

Our networks are composed of streets modeled as lines of nodes by the discretization presented above, see Figure~\ref{fig:paris}. This plays an important role in walker dynamics, therefore we first detail this special case in this section.

\begin{figure}[h!]
\centering
\includegraphics[trim={0cm 0cm 0cm 0cm},clip,
width=\linewidth]{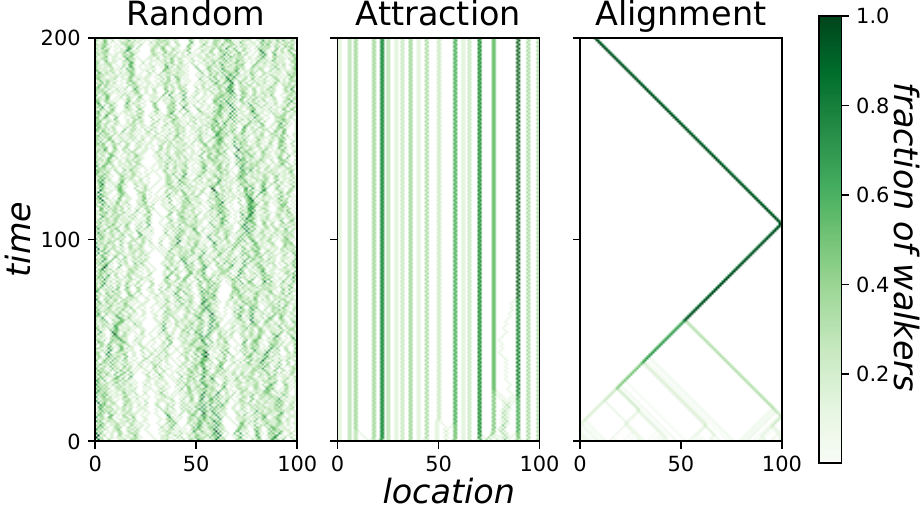}
\caption{Space-time diagrams of three strict tactics on a line using only (from left to right) the Random, Attraction, and Alignment rule. The fraction of walkers is the relative size of groups with respect to the maximum number of walkers in a group in a given run.}
\label{fig:histories}
\end{figure}

Figure~\ref{fig:histories} displays \emph{space-time} diagrams of three different {strict tactics}, {\em i.e.} the ones using only one rule, for $100$ walkers on a line of $100$ nodes. These space-time diagrams display the evolution of walker locations over time. The horizontal axis represents the nodes, numbered from $1$ to $100$. The vertical axis represents time steps, numbered from $1$ to $200$. Then, the color at coordinates $(x,y)$ represents the number of walkers at node $x$ after $y$ time steps, namely $n_x(y)$. The color is white if this number is $0$, and it is the darkest if this number is the maximal value for the considered run. Walkers are initially uniformly distributed on all nodes.

With the Random and with the Attraction strict tactics (Figure~\ref{fig:histories}, left and middle), the largest obtained groups during the run both always contain less than 10 walkers (respectively 6 and 8 on the Figure) which never makes up to 10\% of all walkers. However, the two tactics have distinct behaviors: while the Random one leads to walker constantly moving between groups, the Attraction one leads to stable groups. These groups fail to move along the line, though.

With Alignment (Figure~\ref{fig:histories}, right), a more interesting pattern emerges: all walkers gather into a unique group that achieves flocking since it moves all along the line. %Actually, there are two groups on neighbor nodes in the typical case depicted in the Figure. With respectively 55 and 45 walkers, these groups move together.

In conclusion, the Alignment strict tactic is sufficient to obtain flocking on a street. However, the groups we obtain eventually reach the street extremities, which raises the question of their behavior at street intersections.

\subsection{Extensive exploration}
\label{section:extensive}

In this section, we use an extensive method to explore the wide set of possible tactics on an entire city network:
\begin{itemize}
\item  we consider the Paris street network discretized with parameter $\delta = 10$ meters, leading to a network of $n=130276$ nodes and $m=300736$ links,
\item we consider a set $W$ of $|W|=n$ walkers initially distributed uniformly at random in the network,
\item we perform $1000$ time steps, thus considering reasonably short walks of approximately $10$ kilometers at most,
\item we consider all tactics obtained as combinations of $\alpha_C$ parameter values from $0$ to $1$ by steps of $0.1$ such that $\sum_{C \in \mathcal{C}}\alpha_{C} = 1$,
\item finally, we set $\beta$ large enough to ensure that each walking rule strictly follows its criterion. Indeed, in the following experiments, lowering this value only worsened gathering scores, as it is equivalent to adding noise to walkers choices. %. {\color{red}(we explore less deterministic cases in Section~\ref{sec:dyna})},
\end{itemize}

\begin{figure}[h!]
\centering
    \includegraphics[width=0.8\columnwidth]{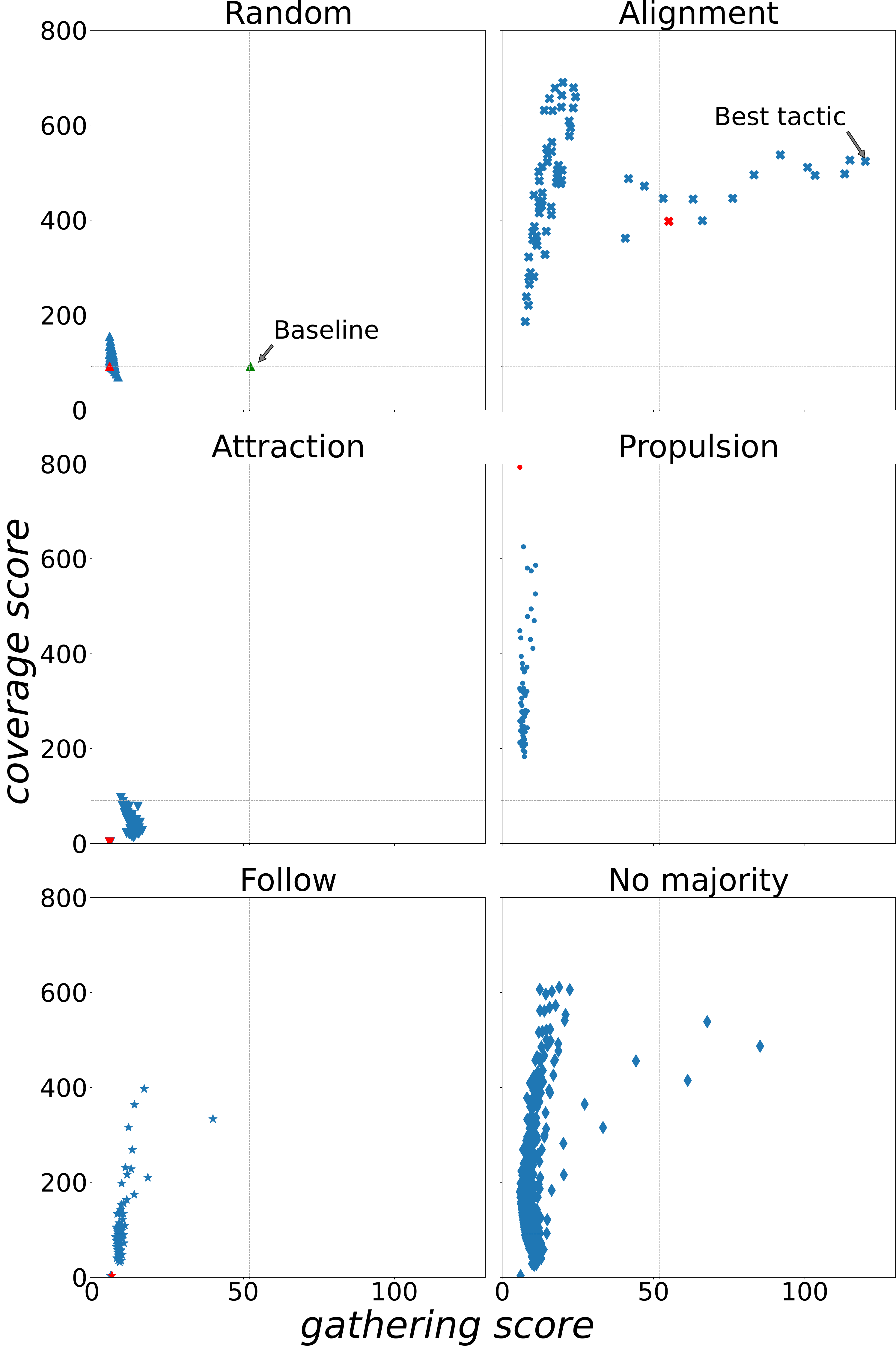}
\caption{\emph{Gathering and coverage scores of all tactics.} Each dot corresponds to the average last step value from ten runs of a tactic. The horizontal axis gives the gathering score, the vertical one gives the coverage score. From left to right and from top to bottom: tactics based mostly on the Random, Alignment, Attraction, Propulsion and Follow rule, respectively. On each of these plots, the red dot is for the strict tactic, that exclusively uses the corresponding rule. The bottom-right plot corresponds to tactics with no prevailing rule.}
\label{fig:scatter}
\end{figure}

%When all agents are on different nodes, $\mathrm{g}(t)=|W|$. If instead all agents are on the same node, it equals $1$.

With this setup, we obtain $1001$ different tactics. We run $10$ simulations of each tactic and plot the average coverage and gathering scores in Figure~\ref{fig:scatter}. In these plots, each dot corresponds to a tactic, defined by a set of $\alpha_C$ parameter values. We split these tactics into six plots: we display a set of tactics on the same plot if they all have $\alpha_C > 0.5$ for the same criterion $C$, and we display on the last plot the set of all other tactics.

We also display in each plot of Figure~\ref{fig:scatter} vertical and a horizontal lines that indicate baseline results. Then, the tactics achieving the best flocking performances are the ones in the upper right corner: they obtain larger and more mobile groups than the baseline.

\begin{figure}[h!]
    \includegraphics[page=1,width=0.98\linewidth]{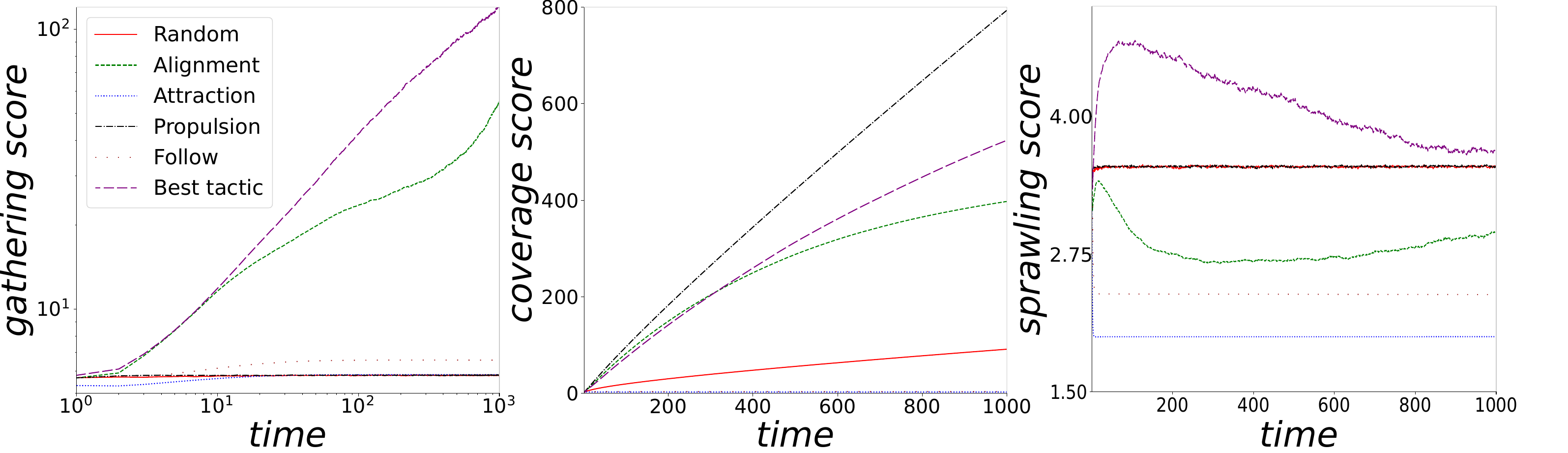}
    \caption{Plots showing the evolution of gathering, coverage and sprawling scores for the strict tactics and the best tactic. Gathering score is in log-log scale for readability.}
    \label{fig:evoltime}
\end{figure}

We pay particular attention to {strict tactics}, which performances are spotted by red %\textcolor{red}{red}
dots in Figure~\ref{fig:scatter}. We also highlight what we identify as the best tactic regarding our scores. It is the tactic with the highest gathering score among tactics that outperform our baseline.
For more insight, we display the evolution of coverage, gathering and sprawling scores over time for these tactics in Figure~\ref{fig:evoltime}.

First notice that all strict tactics have very poor scores, except the Alignment one. This is a consequence of our previous section results, where it was the only one that generates large and mobile groups. Other strict tactics failed to achieve those patterns in a street, and therefore fail on an entire city.

Attraction, Follow and Random strict tactic have almost no dynamic, as walkers do not explore much the network and do not form large groups. Propulsion however displays an excellent coverage score with an almost linear growth on Figure~\ref{fig:evoltime}, as walkers never go back to their previous location.

Note that tactics in the scatter plot (Figure~\ref{fig:scatter}) are not uniformly distributed. Indeed, no tactic has gathering and coverage scores such that its corresponding dot would be in the bottom right corner of the plot. This shows that a high coverage score is a necessary condition to achieve a high gathering score, but it is not sufficient.

Figure~\ref{fig:scatter} clearly shows that Alignment-based tactics (top right plot) outperform others. All other sets of tactics perform poorly, except a few tactics for which no rule weights up more than 50\% (bottom right plot). These tactics actually also use Alignment rule, to a lesser extent. This identifies the Alignment rule as a key building block for flocking tactics.

\subsection{Best tactics}

\definecolor{DarkGreen}{RGB}{1,120,20}

We now focus on the two main tactics that achieve flocking: the Alignment strict tactic and the best tactic (the one that corresponds to the rightmost dot on Figure~\ref{fig:scatter}). Figure~\ref{fig:evoltime} (left and center) displays their scores over time in green %\textcolor{DarkGreen}{green}
and purple %\textcolor{purple}{purple}
colors, respectively.

The plots show that the coverage score of both tactics first rapidly grow, and that this growth significantly decreases over time. Even if these tactics are not as good as the Propulsion strict tactic regarding coverage, they have comparable performances for this metric.

The Propulsion strict tactic has very low gathering scores (Figure~\ref{fig:evoltime}, left), which makes it an irrelevant tactic despite its coverage score. Instead, both the Alignment strict tactic and the best static quickly reach excellent gathering scores. The best tactic significantly outperforms the Alignment strict tactic and has a linear gathering score plot in log-log scale. This means that its evolution has a polynomial growth (of exponent below $1$), indicating a fast growth, but also that the evolution of group size tends to flatten over time. This is due to groups reaching a state where all clusters of groups are in distant regions of the network. Then, it takes longer for groups to meet other groups, merge, and grow in size.

Finally, Figure~\ref{fig:evoltime} (right) displays the sprawling score for all considered tactics. We observe that the best tactic produces a greater sprawling score than Alignment. The sprawling, for those two tactics, is due to groups following each others when they detect another group on a neighbor node, without necessarily merging with it.

With the Alignment strict tactic, the sprawling of groups first very quickly increases, then decreases and stabilizes. This is because this tactic forms groups immediately at the beginning, mostly as lines of walkers following each others. The sprawling reduces as groups reach intersections and split, until the aggregation and splitting dynamics reach an equilibrium.

For the best tactic, groups aggregate into lines for a longer time period, resulting into a much higher sprawling score. It then slowly linearly decreases until the end of the run. This is because the Attraction rule, when chosen in the best tactic, will make the front group wait for the groups behind it, leading to less sprawled clusters.

As explained when defining the metrics, an efficient tactic should have a low sprawling score. The sprawling of the baseline is 1 (up to the third digit), thanks to the collective decision. This is the optimum value.

Our walkers do not have access to collective decision, though. When a cluster of groups arrives at an intersection, at the end of a street, it may split into multiple groups. In the case of a lone group (a cluster with no sprawling) it will split into multiple groups, with equal number of walkers on average. This reduces the gathering score.

However, with a larger sprawling, a cluster keeps most of its groups unchanged after an intersection. Indeed, the first of its groups to reach the intersection node might split into multiple groups, as described above. However, if using an Alignment-based tactic, next walkers will then all follow the largest of the new groups that came from the split. This shows an effective tactic can benefit from a little sprawling to maintain its clusters walker-size.

%This is because walkers only interact if they are on the same node, leading to no sprawling: $\mu_i(t) = 1$ for all $i$. This score only slightly increases when two groups are next to each other by chance. This is the optimum value.

\subsection{Interpretation}
\label{sec:best}

Recall that the best tactic corresponds to the rightmost dot among those in the upper right corner of plots in Figure~\ref{fig:scatter}. It corresponds to the following parameters: $\{ \alpha_{follow}=0.1, \alpha_{align}=0.8, \alpha_{attr} = 0.1 \}$ and a null weight for other rules. This tactic produces groups of $121$ walkers on average, and walkers explore on average $540$ distinct nodes during their $1000$ steps. These scores are more than twice and five times more than what the baseline gets, respectively.

\begin{figure}[h!]
    \captionsetup[subfigure]{labelformat=empty}
    \centering
    \subfloat[1. Pushes forward]{\includegraphics[width=0.4\linewidth]{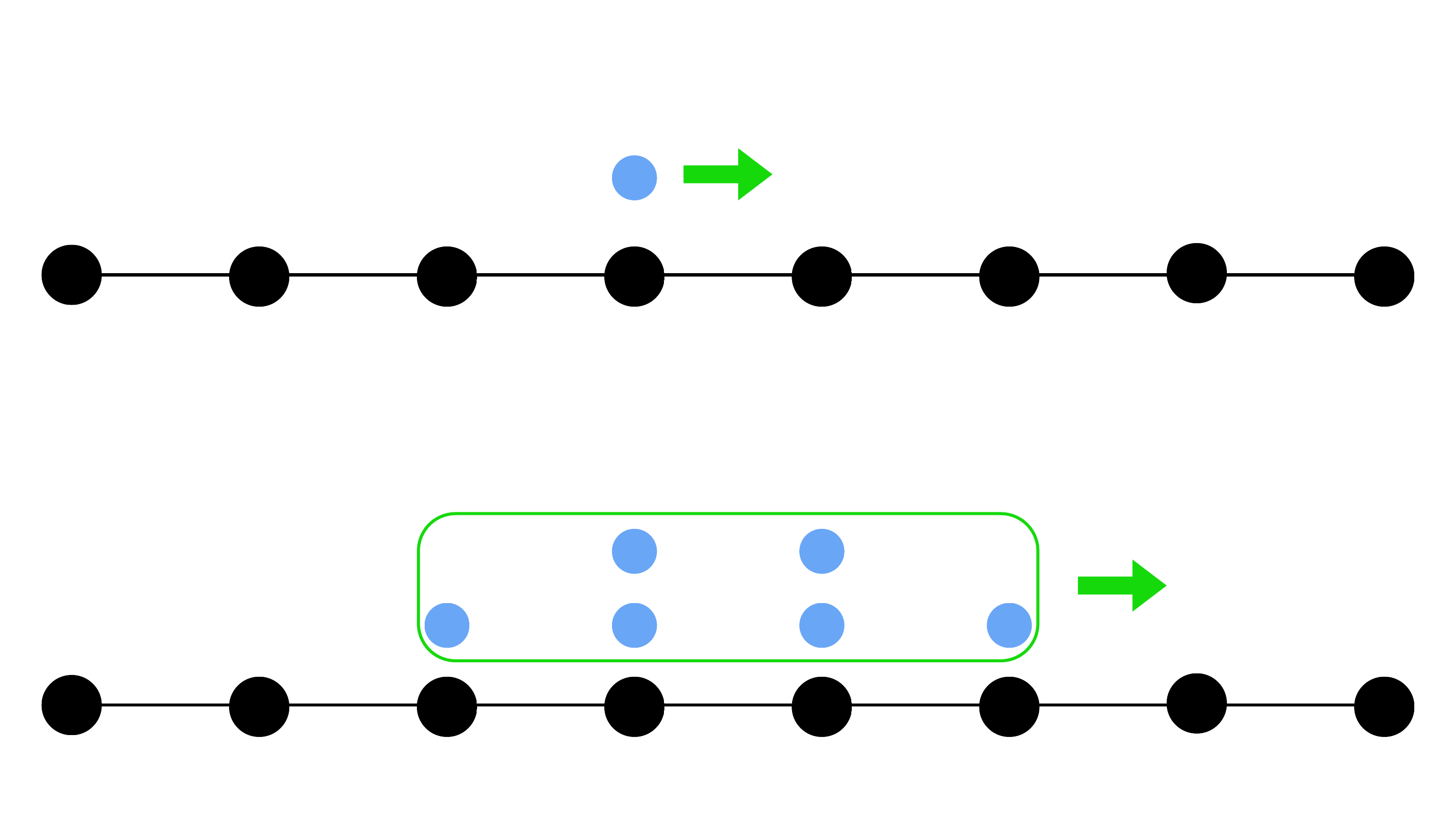}}
    \hspace{20pt}
    \subfloat[2. Merges clusters]{\includegraphics[width=0.4\linewidth]{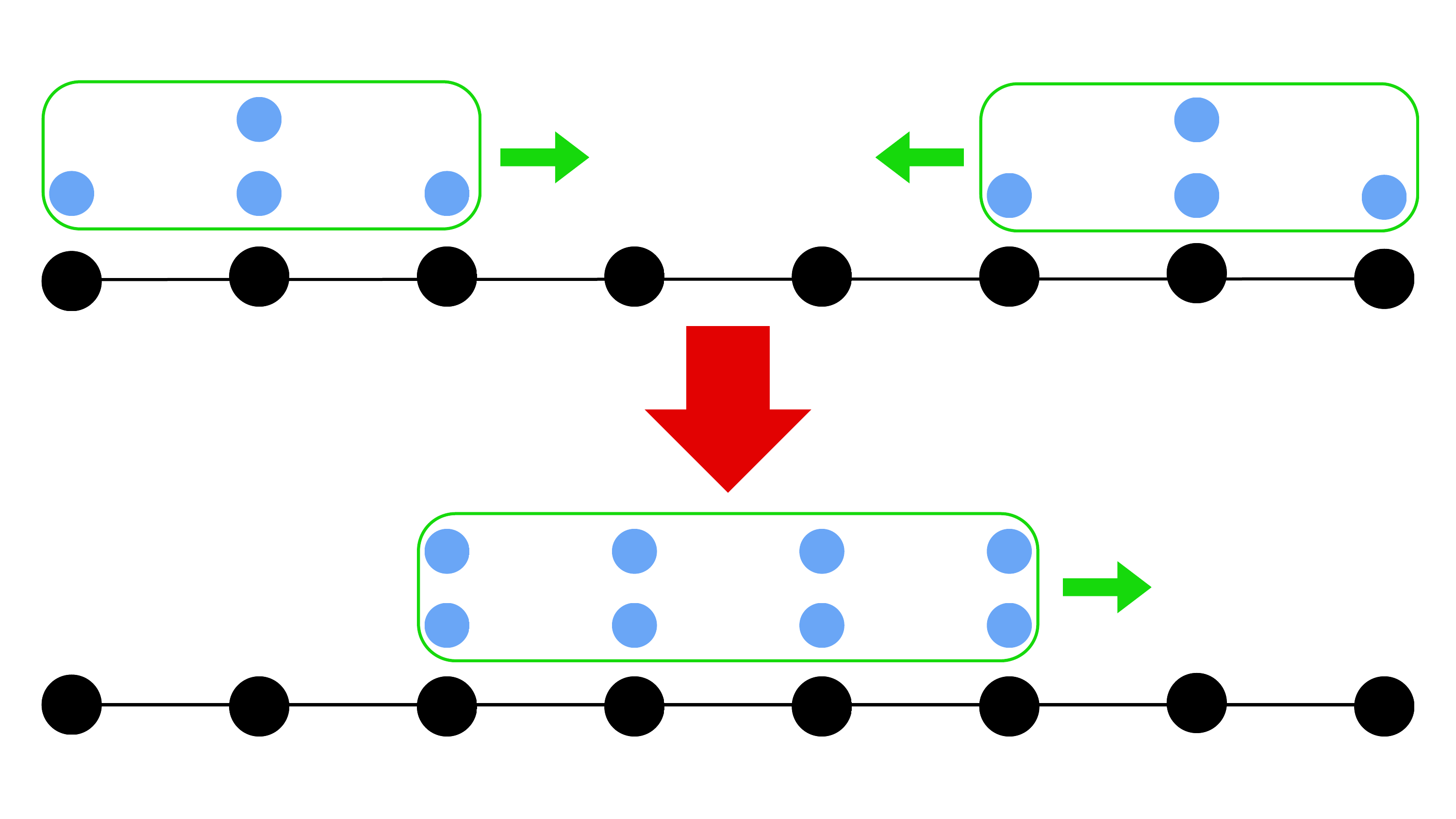}}\\
    \subfloat[3. At intersection...]{\includegraphics[width=0.4\linewidth]{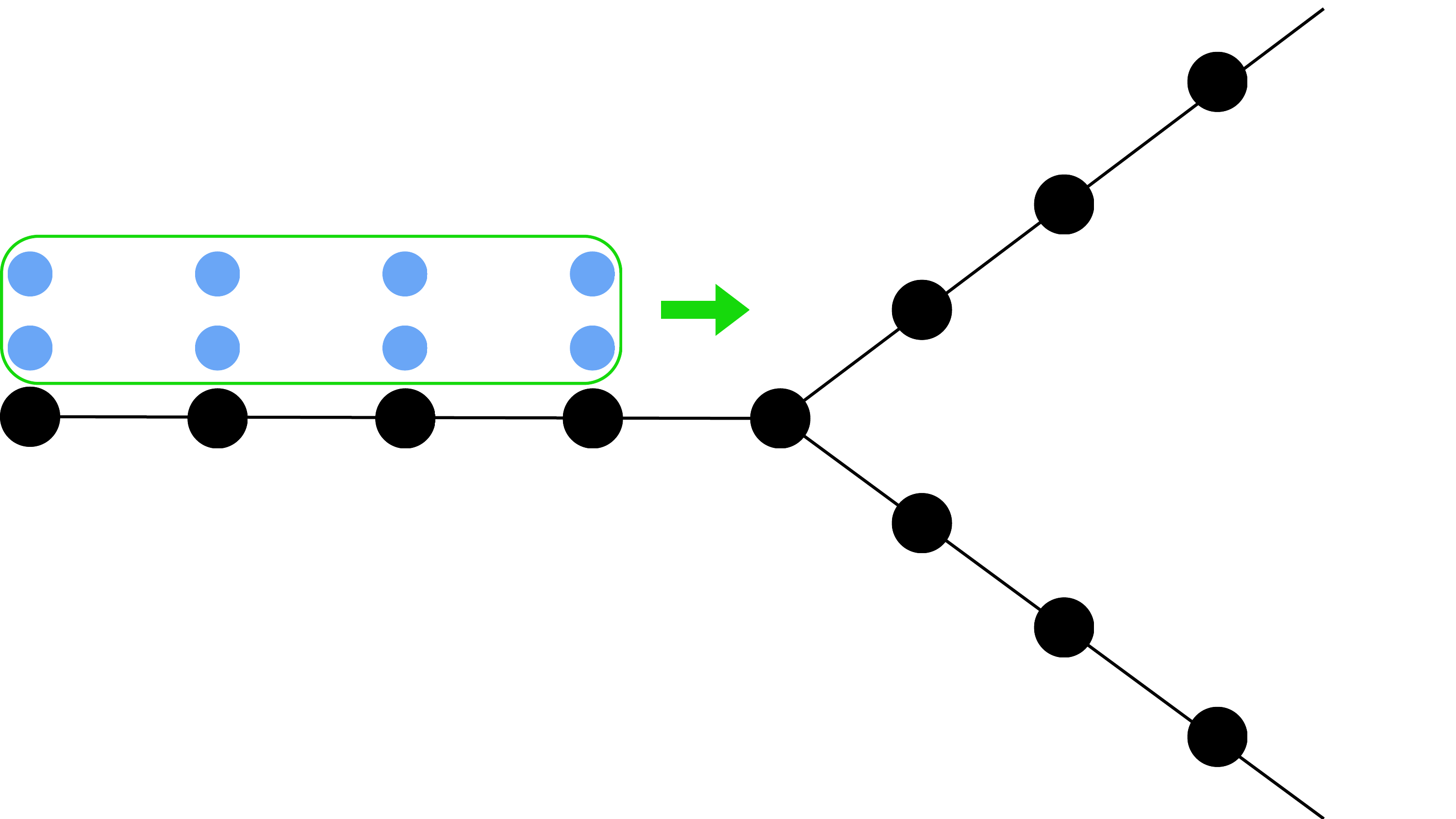}}
    \hspace{20pt}
    \subfloat[... little splitting]{\includegraphics[width=0.4\linewidth]{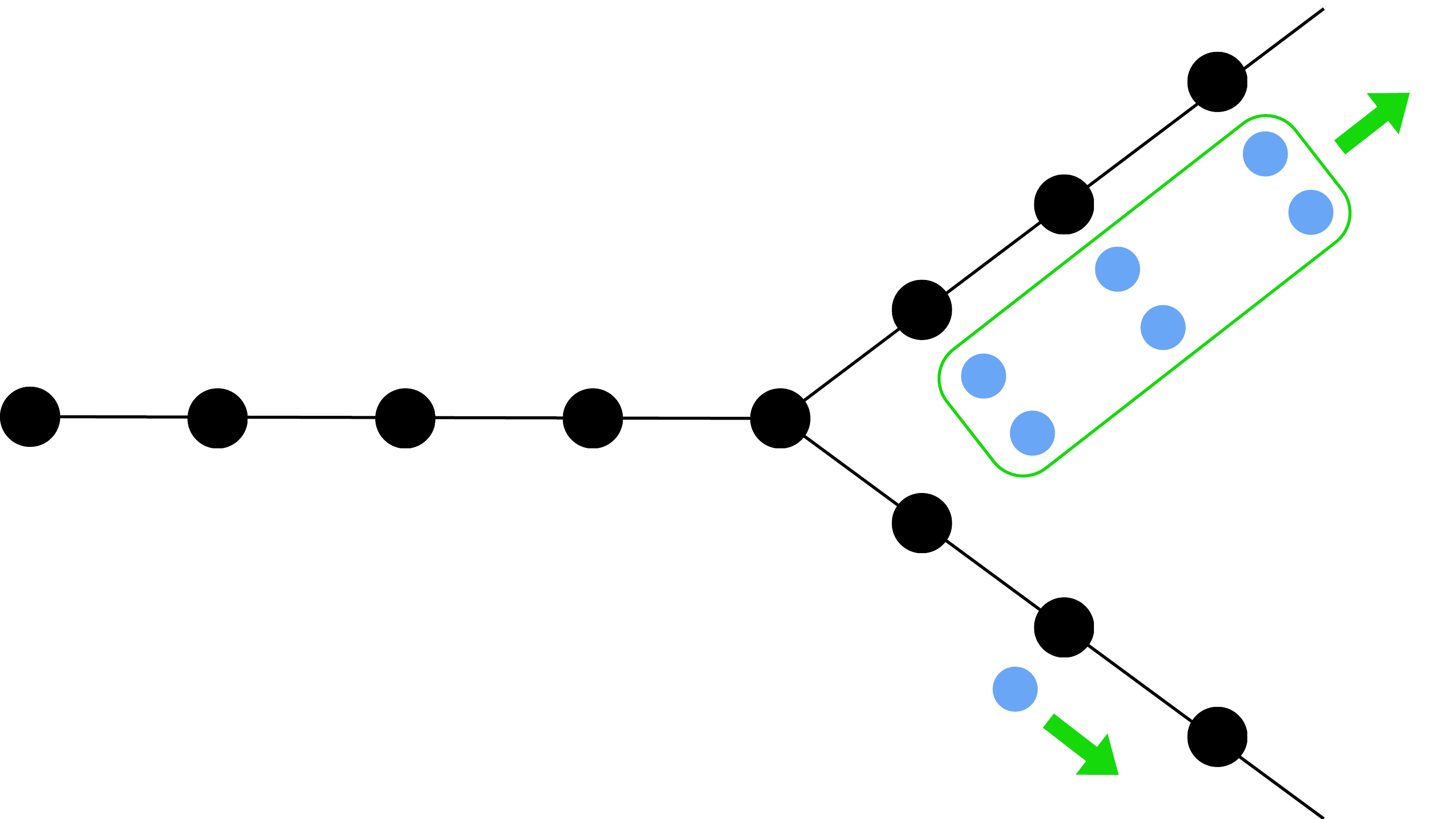}}
    \caption{Typical behaviors of walkers, groups and clusters of groups with the best tactic.}
    \label{fig:drawings}
\end{figure}

Figure~\ref{fig:drawings} illustrates the behavior of this tactic. First, Alignment imposes walkers to move forward, may they be alone or part of a cluster, as shown in the first configuration of Figure~\ref{fig:drawings}. Indeed, a walker $i$ alone at location $x_i(t)= u$ and $x_i(t-1)=v$ will measure a negative flux $\Delta J_{u\rightarrow v}(t) = -1$ at time $t$, while it will be $\Delta J_{u\rightarrow w}(t) = 0$ for all $w\neq v$. This implies the walker never goes back. This effect is left unchanged with multiple walkers in a cluster.

Second, this same rule guarantees that, if two groups cross path, they then merge in a single cluster in which all walkers will follow the same path. Indeed, when a group $u$ cross path with a smaller group $v$, we have $x_i(t+1) = x_j(t)$ and $x_j(t+1) = x_i(t)$ for all $i\in g_u(t)$ and all $j\in g_v(t)$. The net flux is then $\Delta J _{u\rightarrow v}(t) = n_u(t)  -  n_v(t)> 0$ for all walkers, which drive them all in the same direction: the smaller group goes back towards $v$ (where it came from) while the trajectory of the largest group is left unchanged.

However, this rule alone is not perfectly effective at avoiding splitting. The walkers in the first group in a cluster will not always all choose the same node at an intersection, as illustrated in the two bottom pictures of Figure~\ref{fig:drawings}, as the different possible nodes all have a flux equal to~$0$.

In this context, the Attraction rule allows walkers that split in the least chosen direction at an intersection to go backward and avoid loosing sight of the cluster.

Finally, the Follow rule improves the gathering. Indeed, clusters of groups tend to sprawl when walkers use the Alignment rule. In such chain of groups following each others, the Follow rule allows walkers in the front group to move backwards, merging with the group behind them,  while it forces walkers in other groups to move forward to catch up the leading group.

This equilibrium between those three rules gives an outcome where groups flock very efficiently.

Figure~\ref{fig:drawings} also shows why the baseline is not sufficient: it does not imply that walkers move forward at each step. One may try to improve this with a \emph{collective propulsion} (a collective decision with the Propulsion rule instead of the Random one), but such a tactic would not ensure that groups merge when they cross each-other. This is where Alignment rule beats other approaches: walkers merge into a cluster not only if they arrive at the same node, but also if they are neighbors.

\subsection{Group geography}

\begin{figure}[h!]
    \includegraphics[width=0.98\linewidth, clip, trim=25cm 0 0 0]{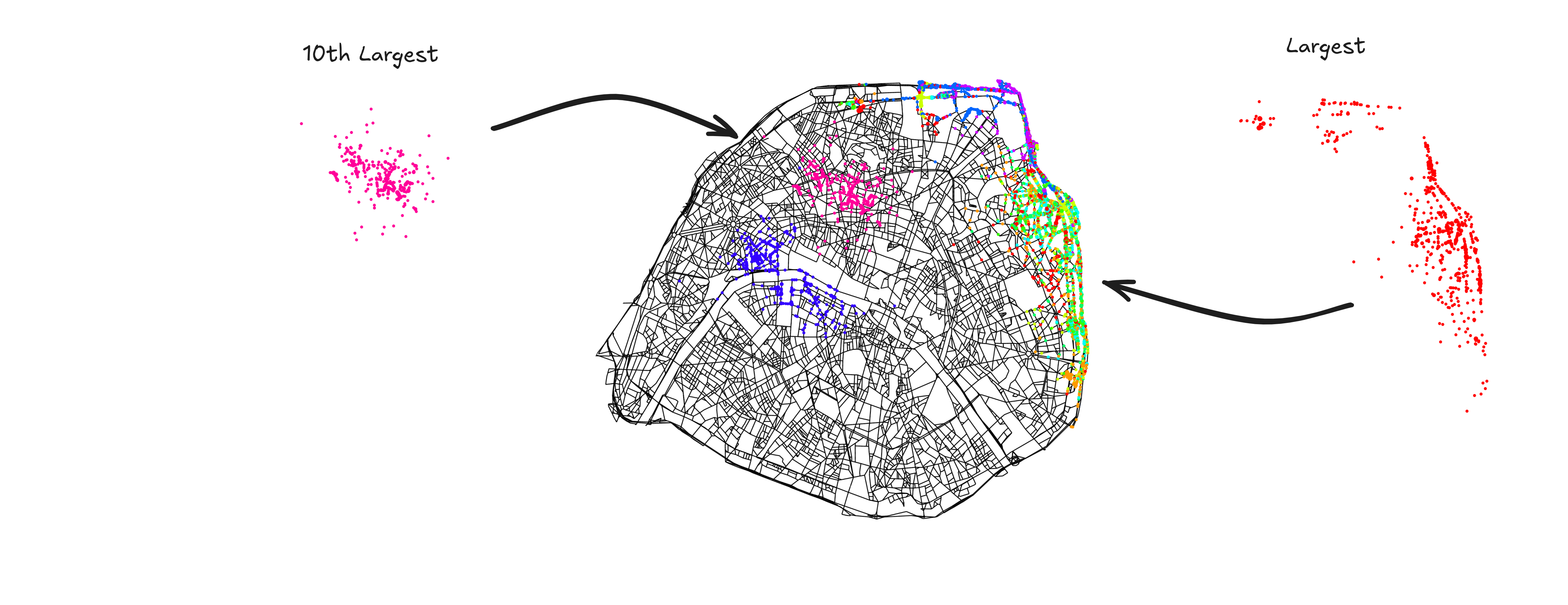}
    \caption{Discretized Paris street network with walkers following the Alignment strict tactic. We color in red the initial position of walkers that end up in the largest group at the end of the run. We do the same, with different colors, for the $10$ largest groups at the end of a run.}
    \label{fig:citywithgroups}
\end{figure}

In order to gain more insight on group formation, we display in Figure~\ref{fig:citywithgroups} the Paris street network for walkers following the Alignment strict tactic. We color nodes according to the group in which a walker that started there ends at the last step of the run. More precisely, we associate a color to each of the ten largest obtained groups. Then, for each walker that ends up in one of these groups, we color its initial location with the group color. The figure presents a typical output observed over many runs.

%with a group color if a walker that started there ends in this group.
%the initial location of the walkers that end in each group with this group color.
Figure~\ref{fig:citywithgroups} clearly shows that the walkers of a group come from a specific and limited region of the network. However these regions are rather wide, and a group does not necessarily gather all walkers that are initially located in the region.

Interestingly, even if two walkers are initially very close, or even neighbors, they often do not end in the same group. As a consequence, the regions defined by groups are entangled, like in the east of Paris (Figure~\ref{fig:citywithgroups}). This is an advantage in our context: it is hard for adversary forces to predict which group a protester will join from its initial position. However, further research is needed to understand what network features lead to such behavior.

\section{Robustness}
\label{section:rob}

It is crucial for protesters to form groups that resist adversary forces that may break them up. In order to explore this robustness, while sticking to our \emph{keep it simple} approach, we do not consider complex models of adversary forces. Instead, we model break ups and the resulting confusion as walkers suddenly following the Random strict tactic for one step. In this way, a group located at a given node splits into smaller groups that move to neighbor nodes, in a way similar to a group of protesters targeted by adversary forces.

More formally, we perform the following experiment: we run the Alignment strict tactic for $100$ steps, then we run the Random strict tactic for one step, and we finally run the Alignment strict tactic again. We perform the same experiment with the best tactic (identified in previous sections) instead of the strict Alignment one.

\begin{figure}[!h]
\centering
    \includegraphics[width=0.98\linewidth]{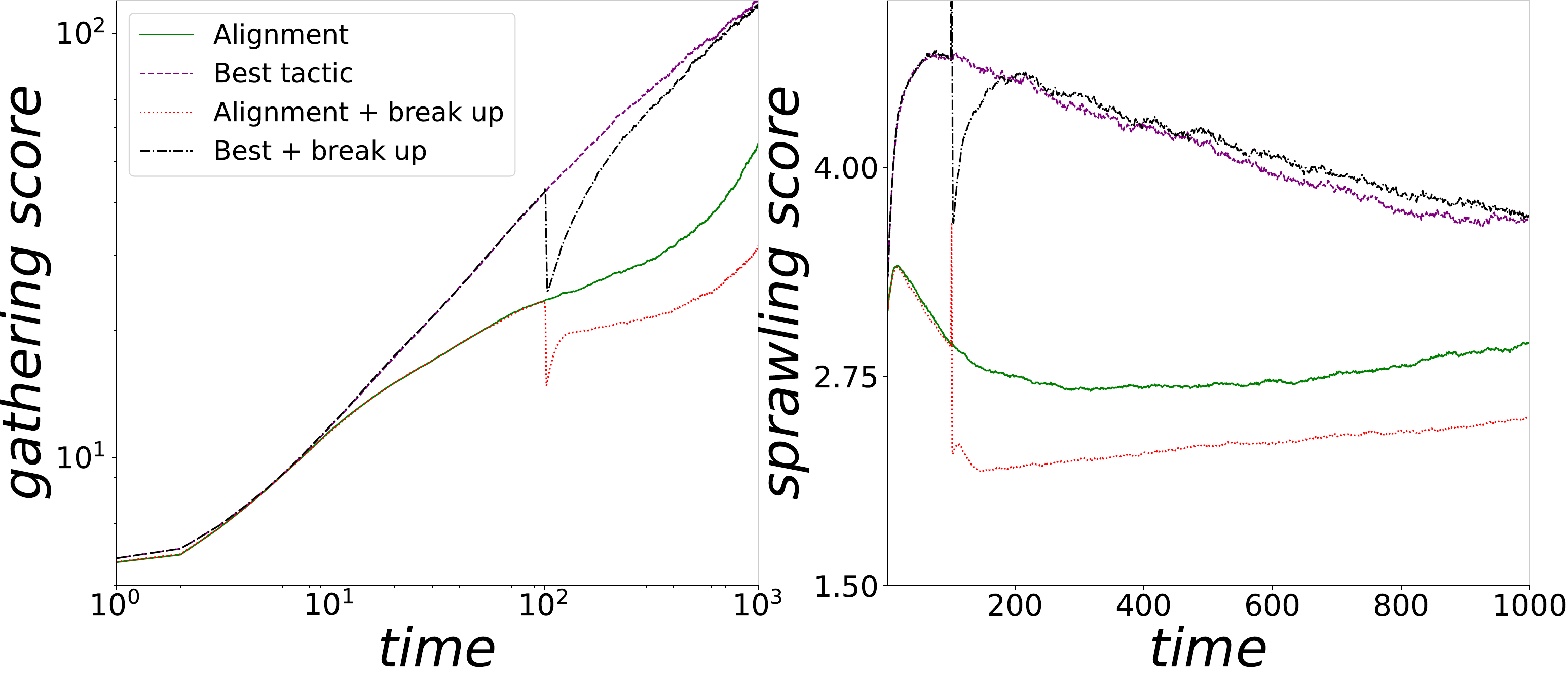}
    \caption{Plots of robustness experiments, similar to Figure~\ref{fig:evoltime}. We display the sprawling and gathering scores for the Alignment strict tactic and the best tactic, as well as for the experiments where all walkers perform a random move at time $100$.}
    \label{fig:dispersion}
\end{figure}

Figure~\ref{fig:dispersion} displays the average observed scores over ten runs of each experiment, together with results of the same experiments without the Random strict tactic step. Since the coverage score is trivially increased by these experiments, which brings no useful insight, we do not display this metric.

The break up simulation clearly appears on both plots. The gathering score plots show that large groups reappear shortly after break ups. The two tactics do not have the same behaviour, though. The strict Alignment tactic needs some time to reach gathering scores close to its pre-break up ones (green %\textcolor{DarkGreen}{green}
curve in Figure~\ref{fig:dispersion}).
Instead, the best tactic is strikingly efficient in quickly recovering from break ups. Indeed, large groups do not only reappear shortly after the break up, they do so even faster than before. This brings its gathering score (\textcolor{black}{black} curve in Figure~\ref{fig:dispersion}) to quickly catch up with the case with no break up (purple %\textcolor{purple}{purple}
curve).

We also observe that, before going back to normal, the sprawling score experiences a sharp increase immediately followed by a sharp decrease. The increase is due to the fact that groups in each cluster first split into several neighbor groups, thus forming a larger cluster. Then, the new groups tend to go forward, due to the Alignment rule, and so they move away from each other, forming distinct clusters. Then, like for the gathering score, the best tactic recovers strikingly well, and faster than the Alignment strict tactic.

%\subsection{Results on different networks and cities}
%\subsection{Variations across cities}
%\label{sec:networks}
%In order to present our model and tactics, we used the specific case of Paris to show our results. We now want to evaluate what differences there are from one city to antoher. Indeed, while city networks hshare commun properties, such as an homogeneous distribution, those networks represent cities that exhibit very different shapes and sizes. We want to see if our tactics are efficient whatever the city. To show this, we will run our simulations with our best tactic on cities of various sizes across the world. We will pay particular attention to the cases of Hong Kong and Seattle, which are cities of the size of Paris, with very different shapes and on which major social movements occured.
\section{Related work}
\label{sec:stateart}

Our work is not concerned with the modeling of actual pedestrian behaviours and trajectory planning \cite{bonnemain2023pedestrians,korbmacher2023time,hoogendoorn2004pedestrian}. Instead, our goal is to design simple tactics that lead to targeted features, like flocking.

Likewise, our work differs from protest models based on thresholds \cite{granovetter1978threshold} or agent-based \cite{epstein2002modeling, kim2011computational, lemos2013agent, agamennone2021riots} approaches. Indeed, these works focus on how people decide to participate in a protest; they do not deal with protester mobility.

Works in \cite{lee2018thin, bu2021extensive} are much closer to ours. Not only their protesters are mobile, but authors also study adversary forces. However, they focus on the strategy these forces should follow to control the crowd, whereas we are more interested in protesters tactics.

Finally, a number of works deal with flocking behaviors of walkers in a variety of situations. We summarize their main features in Table~\ref{tab:rules} and show their limits from our perspective. We discuss them in more details below.

\begin{table}[h!]
\centering
\caption{Important models of collective walks found in literature that provide solutions to problems similar to ours. The $\bullet$ specifies if they succeed at solving one of the issues we deal with.}
\label{tab:rules}
\begin{tabular}{lcccc}
\toprule
\textit{Models} & \textbf{Network} & \textbf{Lattice} & \textbf{Flocking} & \textbf{Fast} \\
\midrule
Random walk \cite{masudaRandomWalksDiffusion2017a} & $\bullet$ & $\bullet$ & & \\
UTS/UXS \cite{Aleliunas1979,Kouck2002}           & $\bullet$ & $\bullet$ & & \\
EAW \cite{yanovski2003distributed}              & $\bullet$ & $\bullet$ & & $\bullet$ \\
SIP \cite{grosskinsky2011condensation}          &          & $\bullet$ & & \\
MIPS \cite{cates2015motility}                   &          & $\bullet$ & & $\bullet$ \\
Boids \cite{reynolds1987flocks}                 &          &          & $\bullet$ & $\bullet$ \\
Cucker-Smale \cite{Cucker2007}                  &          &          & $\bullet$ & $\bullet$ \\
Robot swarm \cite{dorigo2021swarm, Cheraghi2020} &          & $\bullet$ & $\bullet$ & $\bullet$ \\
SVM \cite{vicsek1995novel, raymond2006flocking} &          & $\bullet$ & $\bullet$ & $\bullet$ \\
\textbf{Our work}                                & $\bullet$ & $\bullet$ & $\bullet$ & $\bullet$ \\
\bottomrule
\end{tabular}
\end{table}

\subsection{Walks on networks}

First notice that our baseline %from Section~\ref{section:baseline}
is known as a coalescent random walk~\cite{Cooper2013} and guarantees a non-deterministic gathering. In the field of distributed computing, researchers therefore proposed deterministic solutions for such gathering problems. Walkers then follow a common distributed %deterministic
algorithm to meet on any connected graph~\cite{dessmark2006deterministic, pelc2012deterministic, bhagat2022meet}. The main algorithms are UTS and UXS in Table~\ref{tab:rules}~\cite{Aleliunas1979, Kouck2002}.
%from this community are similar to our baseline, except they are deterministic.
%could be baselines for gathering and flocking.
 %is very similar to the \emph{simple inclusion process} (SIP in Table~\ref{tab:rules})~\cite{grosskinsky2011condensation}.% with condensation but without attraction.

In such works, labeling nodes during walks and knowledge of labels is in general crucial. %are essential components of such protocols. %(even infinite ones).
Moreover the focus is on guarantees regarding long-term gathering. %and this algorithm requires a lot more computation from walkers.
Instead, fast gathering is crucial in our context, as well as very limited memory and computation capacities.

Other works in this line focus on network exploration~\cite{yang2005exploring}, and conclude that backtracking is to avoid, which we indeed observe with the high coverage score of the Propulsion rule.

%We also notably know it is efficient to avoid backtracking in a network~\cite{yang2005exploring} to explore it,

\subsection{Flocking in free space}

Many papers explore flocking since the seminal Boids model~\cite{reynolds1987flocks}. These works typically consider bird swarms or pedestrian crowds that move in \emph{free space}, like the continuous two-dimensional plane. Then, flocking means that a group of agents spontaneously move in the same \emph{direction}~\cite{Cucker2007,OlfatiSaber2006}.

The work of Vicsek~\cite{vicsek1995novel} (SVM in Table~\ref{tab:rules}) notably established that alignment and noise are sufficient to produce flocking in the plan. The combination of alignment and attraction is also an effective way to gather and flock~\cite{henard2023unifying}, as attraction avoids sprawling.

Robot swarms often target objectives very similar to ours~\cite{dorigo2021swarm}. However, they generally use the \emph{local communication model} that exceeds our walkers capabilities. %requires too much for our walkers in term of memory and communication.

Our work differs from all these works by the fact that walkers move in a network, not in free space. Therefore, they do not have a readily usable notion of direction, except in very specific cases where the network is a lattice, for instance. This makes a big difference, and some of our moving criteria may be seen as extensions of free-space concepts to networks.

\subsection{Flocking and networks}

Papers on flocking \emph{and} networks rarely study walks \emph{in} networks. They typically focus on \emph{proximity graphs} of walkers: the graph of which walkers (in free space) are in sight of each other~\cite{olfati2007consensus}. Then, the connectivity of these graphs allows to evaluate if walkers achieve flocking.

Only a few papers are concerned with flocking in networks, but then the considered networks are in general very specific. The motivation typically is to use lattices to approximate free space and study active particles~\cite{cates2015motility}. For instance, the authors of~\cite{raymond2006flocking} explore rules similar to ours in order to study their combinations on a line. They get results similar to the ones we presented in the single street case. %from Section~\ref{sec:streets}.

Finally, some algorithms solve problems close to ours on a network. For instance, \cite{Dereniowski2015} studies the problem of scattering walkers over a graph, when they are initially all gathered at a single node. \emph{Edge ant walk} (EAW in Table~\ref{tab:rules})~\cite{yanovski2003distributed} is a bio-inspired model where walkers communicate by leaving a trace of their passage, useable by other walkers. Such works require capacities that exceed the ones of our walkers.

% To our knowledge, we are the first to explore rules for flocking defined on any network.

\section{Conclusion}
\label{sec:ccl}

This paper proposes a framework for exploring which basic rules are crucial for walkers to gather and flock in street networks. Our findings indicate that the Alignment strict tactic is sufficient for flocking. Combining Alignment with other tactics is even more effective, and obtained groups are very robust to break ups.

This raises many questions on group formation and dynamics over time, and on the role of street network properties in gathering and flocking. For instance, it is unclear why a group emerges at a specific location and gathers walkers from a specific region. Figure~\ref{fig:citywithgroups} is a first step in answering such questions, and it shows they are non-trivial.

Our framework may also be extended in several ways. For instance, the city modelling may be improved with more urban data, like street width or point-of-interest (POI) data (restaurants, stores, or theaters, e.g.). Such information may be used by walkers, and they open the way to studies of the impact of groups on network connectivity.

\medskip\noindent
{\bf Reproducibility.} We publicly provide all code and data used in this paper, our models in C and a Python code to analyse outputs: \url{https://github.com/K-avi/protesting_on_graphs}. We describe how to use it, including how to generate the obtained dataset, on the companion site: \url{https://k-avi.github.io/protesting_on_graphs}.

\section*{Acknowledgements}
\noindent We warmly thank Ivan Mulot-Radojcic for his help with %improving the
implementation and documentation and Nicolas Maudet for his highly valuable feedbacks. This work is funded in part by the MITI CNRS interdisciplinary program.

\bibliographystyle{plain}
\bibliography{flock.bib}

\end{document}